\documentclass[paper]{ieice}
\usepackage[pdftex]{graphicx,xcolor}
\usepackage[fleqn]{amsmath}
\usepackage{newtxtext}
\usepackage[varg]{newtxmath}
\usepackage{booktabs}

\newcommand{\AmSLaTeX}{%
 $\mathcal A$\lower.4ex\hbox{$\!\mathcal M\!$}$\mathcal S$-\LaTeX}
\def\BibTeX{{\rmfamily B\kern-.05em
 \textsc{i\kern-.025em b}\kern-.08em
  T\kern-.1667em\lower.7ex\hbox{E}\kern-.125emX}}
\makeatletter
\def\tmpcite#1{\@ifundefined{b@#1}{\textbf{?}}{\csname b@#1\endcsname}}%
\makeatother

\field{A}
\vol{98}
\no{1}
\title[ ]
      {ISHNE:  Influence Self-attention for Heterogeneous Network Embedding}  
\titlenote{This paper was presented at ...}
\authorlist{%
 \authorentry{Yang Yan}{m}{labelA}\MembershipNumber{1111111}
 \authorentry{Qiuyan Wang}{n}{labelB}[labelC]\MembershipNumber{}
}
\affiliate[labelA]{The author is with the Faculty of Tianjing University of Technology and Education}
\affiliate[labelB]{The author is with the Faculty of Tiangong University}

\begin{document}
\maketitle

\begin{summary}
In recent years, Graph Neural Networks has received enormous attention from academia for its huge potential of modeling the network traits such as macrostructure and single node attributes.
%
%
However, prior mainstream works mainly focus on homogeneous network and lack the capacity to characterize the network heterogeneous property.
Besides, most previous literature cannot model the influence under microscope vision, making it infeasible to model the joint relation between the heterogeneity and mutual interaction within multiple relation type.
In this paper,  we propose an  Influence Self-attention network  to address the difficulties mentioned above.
To model heterogeneity and mutual interaction, we redesign attention mechanism with  influence factor on the single-type relation level, which learns the importance coefficient from its adjacent neighbors under the same meta-path based  patterns .
To incorporate the heterogeneous meta-path in a unified dimension, we developed a self-attention based framework for meta-path relation fusion according to the learned meta-path  coefficient.  
Our experimental results demonstrate that our framework not only achieve higher results than current state-of-the-art baselines, but also show promising vision on depicting heterogeneous interactive relations under complicated network structure.
\end{summary}
\begin{keywords}
 Graph Attention Networks, Heterogeneous Networks, Graph Representation Learning, Passive Influence Relation
\end{keywords}

\section{Introduction}\label{sec:Introduction}
We are now living in a world where all connections come with graph structure, such as traffic network, social network, etc.\
To characterize network topology structure, vertex content and other information, graph embedding representation theory is proposed.  
Graph embedding theory projects the whole graph into a low-dimensional vector space in the form of deep learning paradigm.
Those methodologies include Node2vec~\cite{grover2016node2vec},  DeepWalk~\cite{perozzi2014deepwalk}, Line~\cite{tang2015line}, etc.\

In recent years , one special type of graph called heterogeneous information graph (HIN for short)~\cite{wang2019heterogeneous} has became a hot spot for network mining researchers.
The miscellaneous edges and nodes in heterogeneous graph represent the complex relations such as recommendation system , paper citation network etc.\ 
%
%
%
The most important feature of HIN is the meta path~\cite{dong2017metapath2vec}, which shows the semantic relations in the node-edge tuple.
Taking paper citation network for example,  the relation between two papers can be illustrated as  Paper-Author-Paper (coauthor relations) and Paper-Subject-Paper (peer relations).
From the tuple listed up, we can see the different patterns of connection revel different relation in a heterogeneous graph.
Previous homogeneous graph deep network cannot handle the complex and across-pattern interactions  in the heterogeneous network.

In order to handle such difficulties, various methodology are proposed~\cite{perozzi2014deepwalk}~\cite{dong2017metapath2vec}~\cite{wang2019heterogeneous}.
However, those works still face several limitations.
First, the problem of hidden interactions, such as social influence have not been fully addressed in previous studies, which means valuable social influence of hidden interaction are neglected.
A case in point is the paper citation network.%
If author A has published many influential works,  and B is a follower of A.%
Surely, author A has latent impact on author B in such network.
Second, hidden engagement of adjacent nodes across meta-path relations  have not been discussed in previous studies. 
Although earlier approach has proposed semantic level attention mechanism, their methods cannot address the latent influence issue on ther proposed neural network model.

Based on the previous analysis, we propose a Influence  Self-attention for Heterogeneous Network Embedding framework (denoted as ISHNE) to model the latent influence in the heterogeneous network in this letter.
Our contribution can be summarized as:

(1)To our best knowledge, this is the first attempt to study the latent impact of heterogeneous network based on attention mechanism.
Our work show its promises on crossing meta-path relations in inductive learning where our model achieves better result for the unseen nodes than the SOTA.
(2)To integrate the latent influence relations into graph learning framework, we propose a self-attention model on hierarchical model to fuse the embedding space containing social influence  from different meta-path relations.
(3)We conduct experiments on real heterogeneous graph dataset to evaluate the performance of  ISHNE.
The experimental results show its superiority comparing with the the state-of-the art methodology.
Our method also demonstrates its interruptibility in its result analyze.

\section{Related Works}
%
\subsection{Heterogeneous Information Network}
To tackle the ubiquitous and pervasive multi-modal  interactions in real world,  heterogeneous networks have been proposed and widely used in numerous network mining scenarios.
In HIN~\cite{sun2012mining} , meta-paths are proposed to define the semantic indirect relations in data ming scenarios such as  classification~\cite{ji2010graph}, clustering~\cite{sun2012relation}~\cite{sun2013pathselclus}, recommendation~\cite{chen2017task}~\cite{yu2014personalized} and outlier detection~\cite{gupta2013community}.
elf\subsection{HIN Embedding}
In order to depict the  complex interactions regarding multi-typed links and higher-order meta-path, numerous methodologies are proposed.
Those approaches mainly fall on three categories: proximity-preserving methods, message-passing methods and relation-learning methods.
The goal of proximity-preserving approach is to contain different types of proximity among nodes.
To achieve this, metapath2vec~\cite{perozzi2014deepwalk} proposes a  meta-path guided random walks model to learn the context of node regarding heterogeneous semantics.
Hin2vec~\cite{fu2017hin2vec} considers the possibility of the meta-path between two node s and generate positive and negative tuple on the path generated by random walk algorithm.
PTE~\cite{tang2015pte} decomposes the heterogeneous network into bipartite networks, each of which describes one edge type.
Message-passing methods learn the embedding of attributes via aggregating the information from its adjacent neighbors.
Graph Neural Networks~\cite{kipf2016semi} are widely  used in those approaches.
R-GCN~\cite{schlichtkrull2018modeling} utilizes multiple convolutional layer.
At each convolutional layer, representation vector is updated by accumulating its vectors of neighboring nodes.
HAN~\cite{wang2019heterogeneous}  uses second order  proximity to model meta-path relations. 
The neighbor's weight coefficient are learned by the attention mechanism.
Relation learning methods model each edge as a tuple and design a scoring function.
TransE~\cite{bordes2013translating} designs   translating form of the embedding to minimize the margin-based ranking loss.
ConvE~\cite{dettmers2018convolutional} designs a neural network model to score for the relation tuple. 
In summary, although many aforementioned literature has discussed HIN embedding, the effect of latent influence mechanism has not addressed in previous heterogeneous graph representation learning model.

\section{Preliminaries}
In this section, we will specify  the definition of HIN and our framework.

\textit{Definition 1: Heterogeneous Graph.} 
A heterogeneous graph is defined as a network associated with multiple types of nodes and edges.
Heterogeneous graph can be mathematically  defined as $ G = \{\mathcal{V},  \mathcal{E} \} $, where $\mathcal{V}$ and $\mathcal{E} $ respectively denotes the set of nodes and edges.
Each element of  $ \mathcal{V} $ is associated with  node type function: $ f_{N}:  \mathcal{V}    \rightarrow   \mathcal{A}  $.
Likewise, function $ f_{E} \mathcal{\epsilon} \rightarrow \mathcal{R} $ denotes the link mapping.
Here, $\mathcal{A}$ and $\mathcal{R} $ symbolizes node and link type.

\begin{figure}[h]
	\centering
	\includegraphics[width=1.0\linewidth]{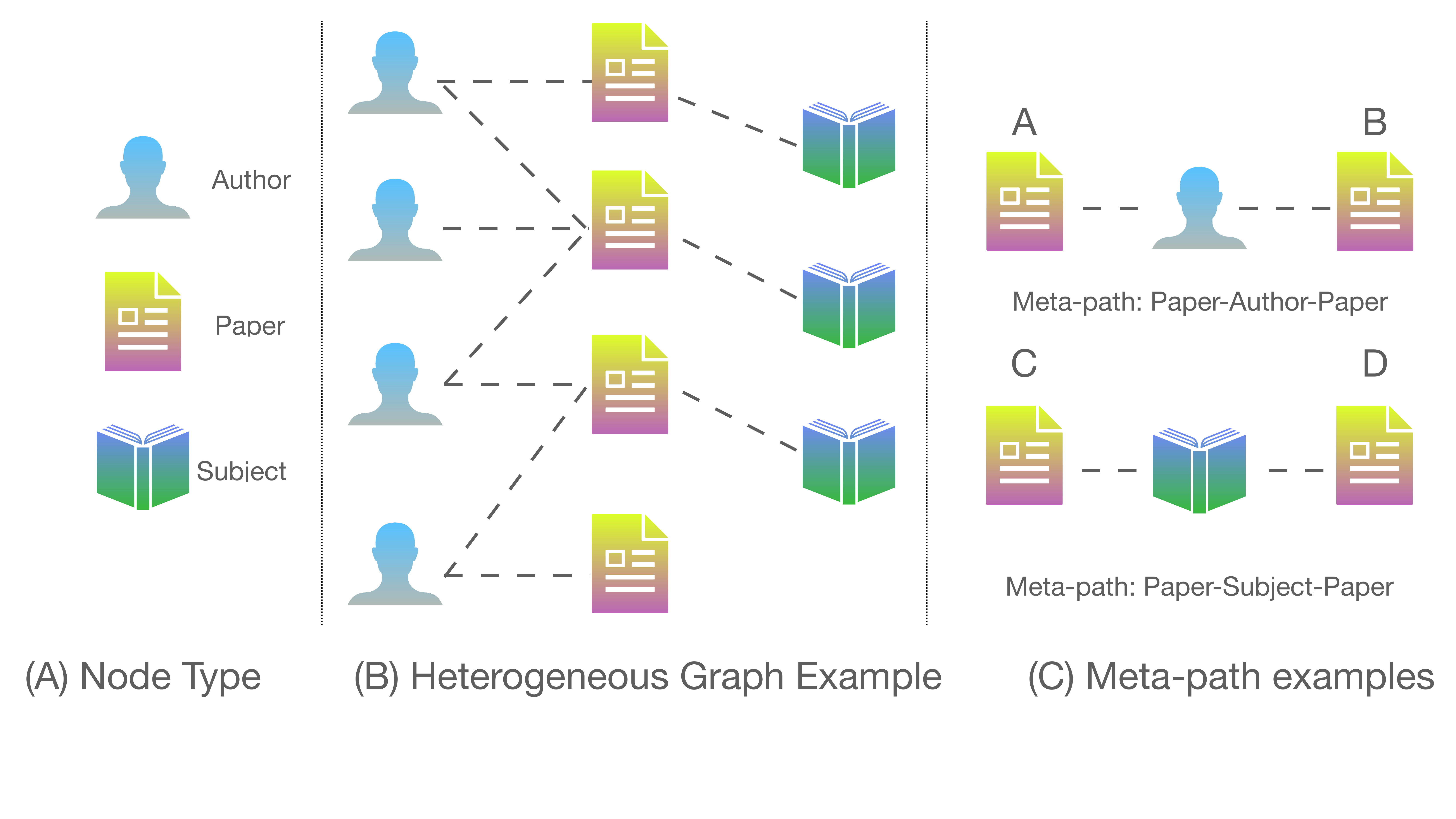}
	\vspace{-1.0cm}
	\caption{Figure~1: An example of heterogeneous graph. Figure~1(A): Three types of nodes in heterogeneous graph. Figure~1(B): Heterogeneous graph consists of multiple nodes and edges. Figure~1(C): Two example of meta-paths.}
	\label{fig:HANExample}
	\vspace{-0.5cm}
\end{figure}

Figure~\ref{fig:HANExample} is an example of heterogeneous  graph to for ACM author-paper-subject relations heterogeneous network.
As shown in Figure~\ref{fig:HANExample}(A) and (B),  there are three different types of nodes (Authors, Papers, Subjects) and two types of linking relations (written-in relation between papers and authors, published-in relation between papers and subjects).

\textit{Definition 2: Meta-path relations.}  A standard form of meta-path relations is $V_1 \stackrel{R_1}{\longrightarrow} V_2 \stackrel{R_2}{\longrightarrow} V_3$, which characterize the relation type $R_1R_2$ series between the vertice $V_1V_2V_3$.

As Figure~\ref{fig:HANExample}(C) illustrated, two papers can be indirectly connected via two different meta-paths: Paper-Author-Paper (denoted as PAP) and Paper-Subject-Paper (denoted as PSP).
In Figure~\ref{fig:HANExample}(C), we can see that node $A$ and node $B$ are connected by meta-path PAP.
Likewise, node $C$ and $D$ are connected by meta-path PSP.
Here, we call $A-B$ and  $C-D$ are meta-path neighbors  on PAP and PSP.
Based on the above preliminary knowledge, we will present a novel neural network, which enables us to exploit the latent influence in the heterogeneous network.
All the symbol we use in this paper are summarized in Table~\ref{table:notationandexplanations}.

\begin{table}

	\caption{Notation and Explanations}
	\tabcolsep 2 pt
	\label{table:notationandexplanations}
	\centering
	\renewcommand{\arraystretch}{1.0}
	\begin{tabular}{c c}
	\toprule[1pt]
	Notation & Explanation \\ \midrule[0.5pt]
	 $  \phi  $ & single meta-path \\
	 $   \textbf{h}  $ & Initial node feature \\
	 $  \textbf{M}_{\phi}  $ & Transformation matrix for type $\phi$  \\
	 $  \textbf{P}_{\phi}$ & influence feature Transformation Matrix \\
	$   \textbf{h}^{p}  $ & Projected influence features vector\\
	
	$e_{ij}^{\Phi}$ & importance coefficient on meta-path $\Phi$ for nodes $i$ and $j$\\
	$\textbf{a}_{\Phi}$ & neighborhood attention vector for meta-path \\
	$a_{ ij }^{ \Phi }$ & Weight coefficient meta-path $\Phi$ for nodes $i$ and $j$  \\
	 $\textbf{N}_i^{\Phi}$ & Meta-path based neighbors\\
	 $\textbf{x}_{\Phi}$ &   Semantic relation embedding	\\
	 $w_{\Phi}$ &   Weight coefficient Semantic relation embedding	\\
	 $\textbf{X}$ &  Final embedding	\\
	\bottomrule[1pt]
	\end{tabular}
	\vspace{-0.5cm}
\end{table}

\section{Our Model}
\begin{figure}[h]
	\centering
	\includegraphics[width=1.0\linewidth]{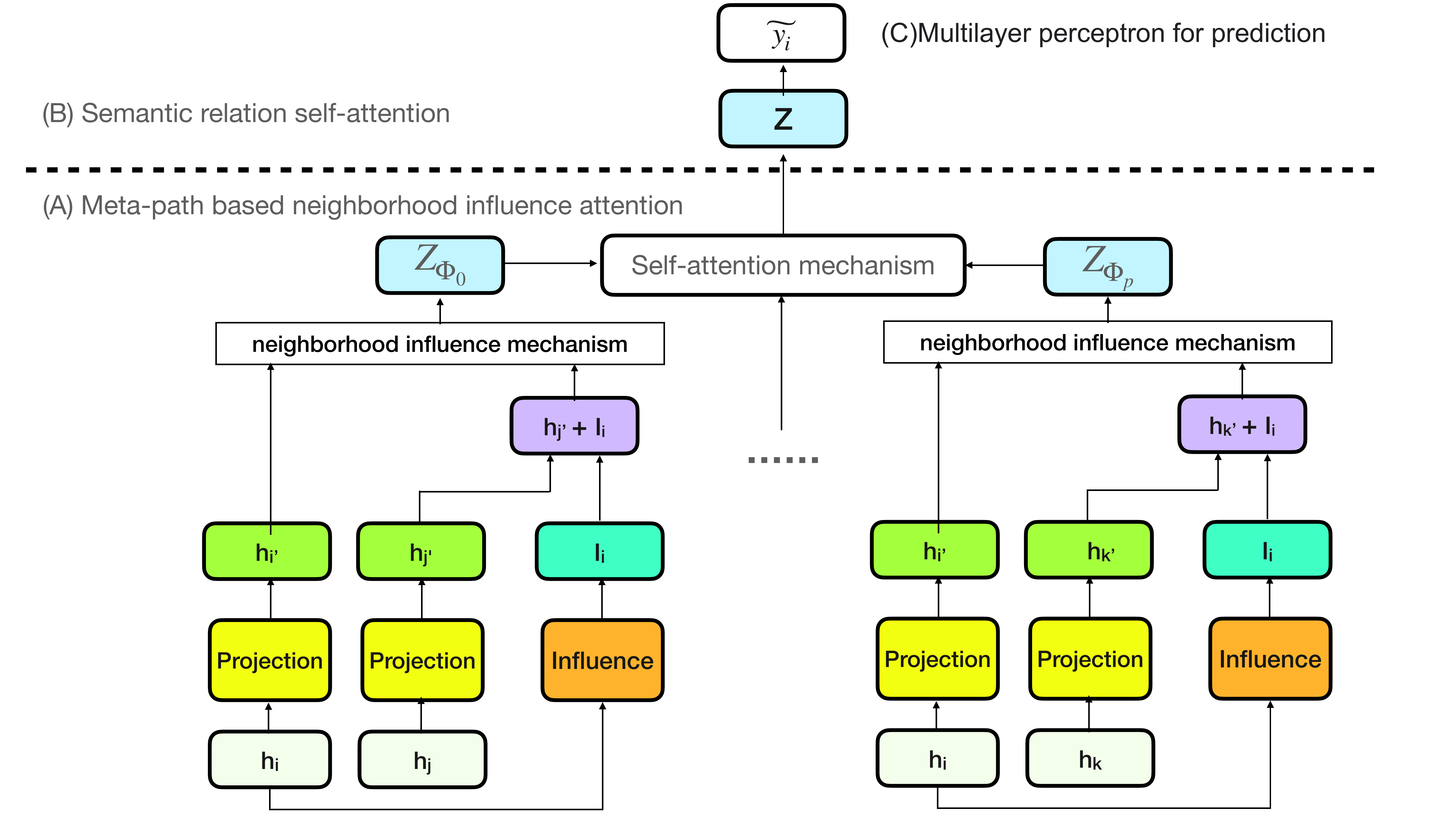}
	\caption{Figure~2: The  framework of ISHNE. (A) Nodes on all meta-paths are projected into  a separate space with its influential feature.(B) Embedding on  meta-paths are merged with self-attention module.(C) Compute the loss in classification task with our ISHNE}
	\label{fig:ishnearechstructure}
	\vspace{-0.5cm}
\end{figure}

In this section, we will describe our proposed   Influence Self-embedding for Heterogeneous Network Embedding framework.
The basic idea of ISHNE is to project every node into different meta-path based embedding space.
To achieve this, local proximities and individual influence of  each node are calculated and integrated by aggregating the meta-path neighborhood information.
The comprehensive embedding of the vertex are fused from all its meta-path embedding by self-attention.
We elaborately design the architecture of ISHNE shown in Figure~\ref{fig:ishnearechstructure}.

\subsection{Meta-Path Based Neighborhood Influence Attention}
Before merging embedding results from each semantic component we should learn that all nodes interacts with its meta-path neighbors. 
Here, we define our neighborhood influence attention model.
For each type of HIN meta path $\phi$, we introduce transformation matrix $M_{\phi}$ to project the node $i$ into the corresponding space of meta-path $\phi$.
The projection function is defined as :
\begin{align}\label{eq:h_i^prime}
\textbf{h}_i^{\prime}  =  \textbf{M}_{\phi}   \cdot  \textbf{h}_i
\end{align}
Here $h_i^{\prime} $ and $h_i$ are projected and original  feature vector of node $i$.
Different from~\cite{wang2019heterogeneous}, we introduce neighborhood influence factor into the attention mechanism.
Specifically, given a node pair $(i,j)$ which is connected by meta-path $\phi$ in academic network.
As we stated in Section~\ref{sec:Introduction}, from the
 perspective of node $j$, node $i$ certainly has some latent effect on him, especially when node $i$ is a well-known researcher.
To character such phenomenon, we impose a influence component on attention computing mechanism.
The influence component is computed as: 
\begin{align}\label{eq:p^influence}
\textbf{h}^p_i =  \textbf{P}_{\phi} \cdot \textbf{h}_i
\end{align}

After that, we leverage the weight coefficient for $(i,j)$ on meta-path $a^{\phi}_{ij}$ in the form of:
\begin{align}\label {eq:a_{ij}^{phi} } 
a_{ij}^{\phi} = \frac{  exp(\sigma( \textbf{a}^T_{\phi} \cdot [ \textbf{h}^{\prime}_i ||  (\textbf{h}^{\prime}_j +  \textbf{h}^p_i)  ] ) )   }{  \sum\limits_{k \in N^{\phi}_i}  exp(\sigma( \textbf{a}^T_{\phi} \cdot [ \textbf{h}^{\prime}_i ||  (\textbf{h}^{\prime}_k +  \textbf{h}^p_i)  ] ) )    }
\end{align}
Here $\sigma$ denotes the activation function.
$| \cdot |$ operator denotes the concatenating operator.
$ \textbf{a}_{\phi}  $ denotes neighborhood influence coefficient on meta path $\phi$. 
On th right part of Equation~\ref{eq:a_{ij}^{phi} },  we add 
$ \textbf{h}^p_i  $ to $ \textbf{h}_j^{\prime} $ to impose the latent influence on node $j$ from node $i$ .
We note that coefficient $ a_{ij}^{} $  is asymmetric because node $i$ and $j$ have different impact to each other.
For node $i$, we will aggregate the feature vector of its meta-path based neighbors in            ~$\textbf{N}_i^{\Phi}$~in the form of Equation~\ref{eq:x_i^{phi}} .
\begin{align}\label{eq:x_i^{phi}}  
\textbf{x}_i^{\phi} = \sigma( \sum\limits_{ j \in N_i^{\phi} }  a_{ij}^{\phi} \cdot \textbf{h}_j^{\prime} )
\end{align}
In order to capture the complicated, rich nature of node features and stabilize the training results, we adopt the multihead attention.
Here, we repeat $K$ times like Equation~\ref{eq:x_iconcatten} to concatenate the embedding vectors as  the input of semantic relation  self-attention model in subsection~\ref{subsec:semanticrelationselfattention}   .
\begin{align}\label{eq:x_iconcatten} 
\textbf{x}_i^{\phi} =   \|_{k=1}^K  \sigma( \sum\limits_{ j \in N_i^{\phi} }  a_{ij}^{\phi} \cdot \textbf{h}_j^{\prime} )   
\end{align}
Given the meta-path set $\{  \Phi_0, \Phi_1,...,\Phi_P \}$, after neighborhood influence attention, we get $P$ group for semantic relation embedding for each node, denoted as $ \textbf{X}_{\phi_0}, \textbf{X}_{\phi_1}, ,...,\textbf{X}_{\phi_p} $.
\subsection{Semantic Relation  Self-Attention Model}\label{subsec:semanticrelationselfattention}
Considering each node can have multiple meta-based relations,
it is necessary to merge the embedding from every meta-path based relations $\phi$.
We here adopt the self-attention mechanism to address the problem of merging the latent influence into a unified semantic space.
To learn the importance of each meta-path,  we device a self-attention based meta-path embedding merging approach.
First, the query matrix $Q$, key matrix $K$ and value matrix  $V$ are computed in Equation~\ref{eq:qvalue}.
\begin{align}\label{eq:qvalue}
\textbf{Q} = \textbf{W}_Q \cdot  \textbf{x}_i^{\phi}  ~~~~~ 
\textbf{K} = \textbf{W}_K  \cdot \textbf{x}_i^{\phi}  ~~~~~
\textbf{V} = \textbf{W}_V \cdot \textbf{x}_i^{\phi}
\end{align}
Next, we leverage the importance of all the semantic relation  self-attention node embedding  which can be explained as the importance of each meta-path. 
For each meta-path $\phi$, the importance of meta path $\textbf{w}_{\phi}$ is computed as Equation~\ref{eq:w_phi_i}.
\begin{align}\label{eq:w_phi_i} 
w_{\phi_i} = \frac{1}{|V|} \sum\limits_{i \in V}      \textbf{q}^T  \cdot  \mathit{softmax}( \frac{\textbf{Q}  \cdot \textbf{K}^T}{\sqrt{d}}  ) \cdot \textbf{V}
\end{align}
Here, $|V|$ denotes the number of nodes in meta-path $\phi$, $d$ is the $2$nd of dimension, $\textbf{q}$ is semantic relation self-attention vector.
The weight $\beta_{\phi_i}$ of meta-path  $\phi_i$ is computed as:
\begin{align}\label{eq:beta_phi} 
\beta_{\phi_i} = \frac{exp(w_{\phi_i})}{\sum\limits_{i=1}^P exp(w_{\phi_i})  }
\end{align}
With the learned weight of each meta-path, we can  obtain the  final result of semantic embedding in Equation~\ref{eq:xfinal}.
\begin{align}\label{eq:xfinal} 
\textbf{X} = \sum\limits_{i=1}^P  \beta_{\phi_i}  \cdot   \textbf{X}_{\phi_i}
\end{align}
The final embedding is applied to a semi-supervised node classification task.
We utilize Cross-Entropy loss function in the task in Equation~\ref{eq:lloss}
\begin{align}\label{eq:lloss}
L= -\sum\lim\limits_{l \in \textbf{y}_l} Y^l ln(C \cdot \textbf{Z})
\end{align}
Here,  $C$ is the parameter of the classifier, $y_l$ is the set of node indices that have labels, $Y^l$ and $X^L$ are the labels of nodes and embedding of labeled nodes respectively. 
\section{Experiments}
We use ACM and IMDB as experimental datasets.
ACM dataset comprises 3025 papers, 5835 authors  and 56 subjects, which constitutes two types of meta-path: $PAP$~(Paper-Author-Paper) and $PSP$~(Paper-Subject-Paper). 
Paper features correspond to elements of a bag-of-words represented of keywords.
IMDB dataset contains 4780 movies, 5841 actors and 2269 directors.
IMDB dataset has three category for the movie nodes: Action, Comedy, Drama.
Meta-paths $MAM$~(Movie-Actor-Movie) and $MDM$~(Movie-Director-Movie) are incorporated in semantic relations.
Movie features employ correspond to elements of a bag-of- words.
We use Macro-F1 and Macro-F1 as evaluation metrics.
We split whole data into three parts.
For ACM, $600$ papers  are for training set.
$300$ papers are for validation set.
The rest are for testing set.
For IMDB, $300$ movies are for training set.
$300$ movies are for validation set.
The rest are for testing set.
\begin{table}
	\caption{Micro-F1 results(\%) Node classification tasks}
	\tabcolsep 2 pt
	\label{table:macrof1tasks}
	\centering
	\renewcommand{\arraystretch}{1.0}
	\begin{tabular}{c c c c c}
		\toprule[1pt]
		Datasets &   GCN & GAT & HAN & ISHNE\\ \midrule[0.5pt]
		ACM  & 75.67 & 26.40 & 82.26 &  \textbf{83.86}\\
		IMDB & 45.78 & 27.99 & 50.58 & \textbf{53.44}\\
		\bottomrule[1pt]
	\end{tabular}
	\vspace{-0.5cm}
\end{table}
\begin{table}
	\caption{Macro-F1 results(\%) Node classification tasks}
	\tabcolsep 2 pt
	\label{table:microf1tasks}
	\centering
	\renewcommand{\arraystretch}{1.0}
	\begin{tabular}{c c c c c}
		\toprule[1pt]
		Datasets &   GCN & GAT & HAN & ISHNE\\ \midrule[0.5pt]
		ACM  & 75.60 & 19.26 & 81.92 & \textbf{83.44}\\
		IMDB & 45.31 & 22.49 & 38.71 & \textbf{51.00}\\
		\bottomrule[1pt]
	\end{tabular}
	\vspace{-0.5cm}
\end{table}

Here, GCN~\cite{kipf2016semi} and GAT~\cite{velivckovic2017graph} are Homogeneous graph embedding approaches.
We perform experiments on all of the meta-paths based homogeneous graphs and pick the best results.
From Table~\ref{table:macrof1tasks} and~\ref{table:microf1tasks}, we can see that our ISHNE achieves $83.86~\% $ and $53.44~\%$ Micro-F1 on ACM and IMDB datasets respectively.
It also achieves $83.44~\%$ and $51.00~\%$ Macro-F1 on two datasets.
We can see the heterogeneous graph methods such as HAN and ISHNE surpass homogeneous graph approaches such as GCN and GAT.
Compared with HAN~\cite{wang2019heterogeneous}, our ISHNE incorporates latent influence in the neighborhood attention computation and fusing meta-path relation process.
Such improvements help us to achieve better results than traditional other attention based mechanism.
We also noticed that on IMDB, ISHNE performs significant superiority than on ACM dataset.
This is mainly because ISHNE is capable of mining rich latent relation on IMDB.

Through the above analysis, we can find that the proposed SHNE achieves the best performance on all datasets. The results show that it is quite important to capture the latent impact of nodes and meta-paths in heterogeneous graph analysis.
\section{Conclusion}
This letter proposes a influence self-attention  network embedding mechanism to tackle the problem of characterizing latent influence relations in heterogeneous graph embedding problem.
Experimental results demonstrate that our approach achieves more satisfying results than SOTA.

\bibliographystyle{ieicetr}
\bibliography{ieice}

\vspace{-2.5cm}
\profile{}{Yang Yan (Ph.D. 2018) received the Ph.D. degree from University of  Chinese Acedemy of Sciences in 2016.  He is currently a lecturer  at School of Information Technology Engineering, Tianjin University of Technology and Education. His research interests include Social Network, Data mining, coding theory and cryptography.}
\label{profile}

\profile{}{Qiuyan Wang (Ph.D. 2016) received the Ph.D. degree from University of  Chinese Acedemy of Sciences in
	2016. She is currently a lecturer  at School of Computer Sciences and Technology, Tiangong University. Her
	research interests include coding theory and cryptography.}

\end{document}